\documentclass[12pt]{article}
\usepackage[dvips]{graphicx}
\usepackage{subfigure}
\renewcommand{\baselinestretch}{2}

\begin{document}

\renewcommand{\thefootnote}{\fnsymbol{footnote}}
\title{Cooperativity and Stability \\ in a Langevin Model of Protein
Folding\thanks{This paper has been submitted for publication to the
Journal of Chemical Physics}}

\author{Gabriel F. Berriz, Alexander M. Gutin, and Eugene I.
Shakhnovich\thanks{Corresponding author.}}
\renewcommand{\thefootnote}{\arabic{footnote}}
\date{\today}
\maketitle
\begin{center}
Harvard University, Department of Chemistry\\
12 Oxford Street, Cambridge MA 02138
\end{center}
\newpage
\begin{abstract}
We present two simplified models of protein dynamics based on
Langevin's equation of motion in a viscous medium.  We explore the
effect of the potential energy function's symmetry on the kinetics and
thermodynamics of simulated folding.  We find that an isotropic
potential energy function produces, at best, a modest degree of
cooperativity.  In contrast, a suitable anisotropic potential energy
function delivers strong cooperativity.

\end{abstract}
\newpage

\section*{Introduction}

Physical chemists have long marvelled at the ability of globular
proteins to fold quickly from a random denatured state to a specific
three-dimensional conformation.  Though much has been learned about
this phenomenon since then, we still lack a satisfactory theory of the
folding process.

Theoretical work in this area has, from the beginning, relied heavily
on computer simulations.  The early efforts to model protein folding
numerically attempted to represent real proteins for which
crystallographic and other experimental data were available.  This
type of simulation study, which is still much used today, starts with
as realistic a representation of the modelled protein and the physical
interactions between its constituents as is computationally
affordable.  We may describe this as a ``top-down'' approach.  In
contrast, a more recent approach proceeds from the bottom up.  It
starts with the simplest model that still bears a minimal resemblance
to a protein while still being complex enough to pose non-trivial
theoretical questions.  The most important examples of this strategy
are the lattice models of protein folding (e.g. \cite{ssk-nature94,
chan-dill94, shakh94.0, hao-scheraga94, socci-onuchic94}).  These
models reduce the protein to a linear self-avoiding chain of beads
that are constrained to inhabit points in a discrete three-dimensional
lattice; the kinetics and the thermodynamics features of the models
are accordingly streamlined.  These very simple lattice models have
allowed researchers to tackle theoretical questions that are
computationally prohibitive to top-down models.

Despite their great usefulness, however, the extreme simplicity of
lattice models gives rise to the legitimate question of how much the
results obtained with lattice models depend on the severe geometric
constraints imposed by the lattice.  Moreover, these very constraints
preclude the investigation of important questions that are too
sensitive to finer geometric details than can be represented in
discretized space.  Therefore, both to independently test the results
obtained by lattice models, and to address questions sensitive to fine
geometrical details, it would be useful to have simple off-lattice
models of protein folding.  A few investigators have produced such
models, and obtained encouraging results
(e.g. \cite{honeycutt-thirumalai90,honeycutt-thirumalai92,grubmuller-tavan94,gron-doniach94}).  One
aspect that these previous studies have not fully addressed, however,
is the degree of cooperativity observed in the unfolded
$\leftrightarrow$ folded transition.  This question is the focus of
the present work.

\section*{Model}

We model proteins as linear self-avoiding chains of hard spheres
(corresponding roughly to amino acids) connected by rigid rods
(corresponding roughly to $C_{\alpha}-C_{\alpha}$ pseudobonds).
Except for the excluded-volume constraints imposed by the hard
spheres, the chain is freely jointed at each monomer.

We use the basic strategy of classical molecular dynamics algorithms:
at each time step, each monomer is displaced slightly, according to
the forces acting on it.  The net force on each monomer is the sum of
a ``regular'' component, derived from the potential energy field
generated by nearby monomers, and a ``random'' or ``noise'' component,
drawn randomly from a Maxwell distribution.  Our displacements are
computed according to Langevin's equation of motion in a viscous
medium.  (For more details on the use of Langevin dynamics to model
protein motion, consult~\cite{gron-doniach94} and
~\cite{honeycutt-thirumalai92}.  Many aspects of our algorithm derive
from the one presented in~\cite{gron-doniach94}.  One important
difference is our use of a SHAKE algorithm to enforce bond-length
constraints, thereby avoiding the bond-length divergence as $t
\rightarrow \infty$ reported in that work.)

For any chain configuration, the potential energy $U$ is computed as

\begin{equation}
U = \sum_{i = 1}^{N-2} \sum_{j = i + 2}^{N} U_{ij} ,
\label{totalPotentialEnergy}
\end{equation}

where $U_{ij}$ is the potential energy of the pair $i j$ of monomers.

In addition to the hard-sphere repulsion between all pairs of
monomers, there is also a short-range attraction between those
monomers that are in contact in the native conformation.  (The precise
definition of a native contact is given below.)  The strength of this
attraction is the same for all such pairs.  (This aspect of the model
is entirely analogous to that proposed by Go et
al. in~\cite{taketomi-ueda-go75}, and hence we refer to it as ``the Go
prescription.''  This class of models may be viewed as the {\em
limiting case} for optimal design in more realistic ``sequence''
models, i.e. models in which the pairwise energies of interaction
depend on the chemical identity of the interacting residues.)  Hence,
the potential energy function, in general, consists of a repulsive
part and an attractive part, the latter being zero for all but those
monomer pairs that form native contacts.  The two variants of our
model presented in this paper differ solely in the function used to
represent the attractive part of the potential acting at native
contacts.  In one model, this attractive potential is spherically
symmetric, whereas in the other it is completely asymmetric.  In this
paper, we will refer to them as the ``isotropic'' and ``anisotropic''
models, respectively.

We describe the potential energy function used in the isotropic model
via the expression for the terms $U_{ij}$ in
Equation~\ref{totalPotentialEnergy}:

\begin{equation}
U_{ij}(r) = U_{0} \left[\left(\frac{r_0}{r}\right)^{12}
- 2 \Delta_{ij}\left(\frac{r_0}{r}\right)^6 \right]
\label{isoPotential0}
\end{equation}

Except for the term $\Delta_{ij}$ on the right, this is a standard
Lennard-Jones 6-12 potential.  We define $\Delta_{ij}$ to be 1 if
monomers $i$ and $j$ are within a distance of 2 bond-lengths of each
other in the native structure (i.e. they constitute a {\em native
contact} for the purpose of Go's prescription); otherwise $\Delta_{ij}
= 0$.

\begin{figure}
\begin{picture}(6,10)
\put(4.2,0.5){${\bf \hat{r}}$}
\put(3.3,4.5){${\bf \hat{\Theta}}$}
\put(4.7,5.3){${\bf \hat{u}}_{ij}$}
\put(6.7,5.5){$r$}
\put(9.6,5.6){${\bf \hat{u}}_{ji}$}
\large
\put(8.4,5.3){$\theta$}
\put(7.7,2.1){$i$}
\put(11,7.6){$j$}
\normalsize
\scalebox{0.85}{\includegraphics[bb=-0.5in 0.25in 4.5in 5in, clip=true]{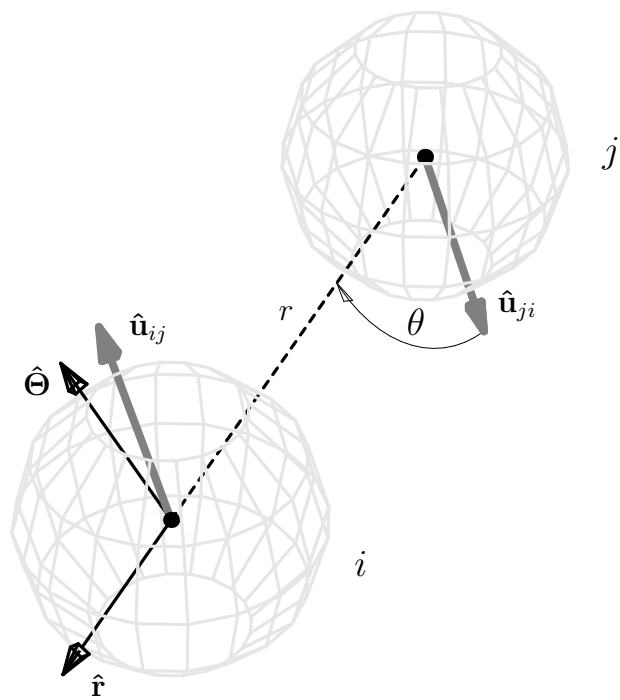}}
\end{picture}
\renewcommand{\baselinestretch}{1}
\caption{Definition of the $\theta$ angle, used in
Equation~\ref{anisoPotential0}.}
\label{thetaDef}
\end{figure}

To write the corresponding expression for the anisotropic model we
must first define one additional variable, namely the angle $\theta$
shown in Figure~\ref{thetaDef}.  Starting from an {\em arbitrary}
Cartesian representation of the native structure, for each pair $i,j$
of monomers that form a contact in this reference native structure
(using the distance criterion given in the previous paragraph), we
define a unit vector ${\bf \hat{u}}_{ij}$ pointing from monomer $i$ to
monomer $j$, and, conversely, another unit vector ${\bf
\hat{u}}_{ji} = - {\bf \hat{u}}_{ij}$ in the opposite direction.  For
the sake of computational efficiency, in the version of our
anisotropic model presented here, once the orientations of these unit
vectors are defined at the start of the simulation, they {\em remain
constant} throughout.  (In a more general model, these vectors would
be free to rotate, as long as all the unit vectors emanating from a
given monomer rotated together as a rigid body, i.e. preserving all
the angles between them; for further remarks on this important aspect
of the model see the Discussion section.)  Now, we define the angle
$\theta \in [0,\pi]$ as that between the unit vector ${\bf
\hat{u}}_{ji}$ (or, equivalently, ${\bf \hat{u}}_{ij}$) and the line
connecting monomers $i$ and $j$ (see Figure~\ref{thetaDef}).  It
measures the deviation of the $ij$ couple's (signed) orientation from
what it is in the native state.  For the sake of completeness, if
monomers $i$ and $j$ do not participate in a native contact, we define
$\theta = 0$.  Now we can write the expression analogous to
Equation~\ref{isoPotential0} for the anisotropic case:

\begin{equation}
U_{ij}(r, \theta) = U_0 \left[\left(\frac{r_0}{r}\right)^{12}
- 2 \Delta_{ij}\left(\frac{r_0}{r}\right)^6 e^{- 6 \theta^2} \right].
\label{anisoPotential0}
\end{equation}

The new element in this expression is the test function $e^{-6
\theta^2}$ used to penalize deviations of $\theta$ from
zero~\footnote{Our choice for this function is quite arbitrary.  In
principle, our purposes only require that 1) it have value 1 at
$\theta = 0$, and decay monotonically to 0 as as $\theta \rightarrow
\pi$; and 2) it be differentiable almost everywhere in the
interval $[0,\pi]$ (since we must be able to compute the gradient of
the potential to get the resulting regular force).  The function we
chose decays rapidly to 0 as $\theta \rightarrow \pi$, but we have not
studied the question of how steep this decay needs to be to yield the
results we report here.}.

\section*{Methods}

The overall structure of the algorithm we use is standard: at every
time step, random forces are generated and regular forces computed;
from these, the unconstrained displacements are computed; finally, we
use a SHAKE~\cite{shake0,shake1} subroutine to enforce bond-length
constraints.

The 3 spatial components of the random force ${\bf F}_{rand, i}$ on
monomer $i$ are independently generated as normally distributed random
variables with zero mean and variance $2D/\Delta t = 2k_BT/\gamma\Delta
t$, where $D = k_BT/\gamma$ is the diffusion coefficient, $k_B$ the
Boltzmann constant, $T$ the temperature, $\gamma$ the frictional drag
coefficient, and $\Delta t$ the simulation's time-step.  This choice
of variance for the forces ensures that, in the absence of regular
forces, the variance of the resulting displacement will be $6 D \Delta
t = 3 \times 2 k_BT\Delta t/\gamma$.  We rescale to dimensionless
parameters, setting $k_B = 1$ and $\gamma = 1$; therefore, the diffusion
coefficient becomes numerically equal to the temperature.  With these
definitions, one time unit corresponds to the time it takes, at $T =
1$, for a particle to diffuse, on average, a distance of $\sqrt{6}
\approx 2.45$ bond lengths.  In our simulations we use a time step
$\Delta t$ never greater than 0.001 time units.

The contribution to the regular force produced by monomer $j$ on
monomer $i$ is computed as $-\nabla U_{ij}$.  For the following
exposition, we rewrite Equations~\ref{isoPotential0} and~\ref{anisoPotential0} as

\begin{equation}
U_{ij}(r, \theta) = U_0 \left[\left(\frac{r_0}{r}\right)^{12}
- 2 \Delta_{ij}\left(\frac{r_0}{r}\right)^6 A(\theta) \right]
\label{generalPotential}
\end{equation}

(where now equations~\ref{isoPotential0} and~\ref{anisoPotential0} may
be recovered from equation~\ref{generalPotential} by setting
$A(\theta) = 1$ and $A(\theta) = e^{-6
\theta^2}$, respectively).  We set $U_0 = 1$, and therefore,
$U_0$ may be regarded as our energy unit.  Likewise, we use the bond
length as our basic unit of distance (although occasionally we use the
conversion 1 bond length = $C_{\alpha}-C_{\alpha}$ distance = 3.8~\AA,
to make our results more directly comparable to experimental values).
To ensure that the chain does not cross itself, we make such crossings
energetically prohibitive by setting $r_0 = 1.5$ bond lengths.

Evaluating $-\nabla U_{ij}$ in local polar coordinates at
$(r,\theta)$,
we get

\begin{eqnarray}
{\bf F}_{ij} &=& -\left(\frac{\partial U_{ij}}{\partial r}\,{\bf \hat{r}}
+\frac{1}{r} \,\frac{\partial U_{ij}}{\partial \theta}\,
{\bf\hat{\mbox{\boldmath $\Theta$}}}
\right)\nonumber \\ &=& 12\; U_{0}\left(\frac{r_{0}^{12}}{r^{13}} -
\frac{\Delta_{ij} r_{0}^{6}}{r^7}A(\theta)\right) {\bf
\hat{r}}
+ \frac{2 U_{0}\Delta_{ij} r_{0}^{6}}{r^7}
A\,'(\theta){\bf\hat{\mbox{\boldmath $\Theta$}}}
\nonumber,
\end{eqnarray}

where the terms $r, \theta, {\bf \hat{r}}$, and ${\bf\hat{\Theta}}$
are as shown in Figure~\ref{thetaDef}.  In particular, ${\bf
\hat{r}}$, and ${\bf\hat{\Theta}}$ are orthogonal unit vectors.  The
latter is coplanar with ${\bf \hat{u}}_{ji}$ and ${\bf \hat{r}}$.  (Of
the two directions that ${\bf\hat{\Theta}}$ can have, and still be
both orthogonal to ${\bf \hat{r}}$ and coplanar with ${\bf
\hat{u}}_{ji}$ and ${\bf \hat{r}}$, we have chosen the one along which
moving monomer $i$ increases $\theta$; see Figure~\ref{thetaDef}).

Note that in the isotropic case, $A\,'$ is identically zero, so the
regular force consists of a radial component only.  In
the anisotropic case, on the other hand, $A\,'(0) = 0$, and
$A\,'(\theta) < 0,\, \forall \,\theta \in (0,\pi)$.

The total regular force ${\bf F}_{reg, i}$ on monomer $i$ is simply

\[ {\bf F}_{reg, i} = \sum_{j \ni |j - i| \geq 2}^{N}{\bf F}_{ij}, \]

(In fact, for the sake of computational speed, the algorithm ignores
the contributions to the regular force from monomers outside of a
cutoff radius $r_{\mbox{\scriptsize cutoff}} = 1.8 \, r_0 = 2.7$.)

We obtain the total force ${\bf F}_{i}$ on monomer $i$ from

\[ {\bf F}_{i} = {\bf F}_{rand,i} + {\bf F}_{reg,i}, \]

and the corresponding (unconstrained) displacement $\Delta {\bf
R}_{i}$ from

\[ \Delta {\bf R}_{i} = \Delta t \; {\bf F}_{i}, \]

The above expression comes from the Langevin
equation~\cite{mcquarrie_statmech}:

\[ \gamma {\bf v}_{i} + m_i \frac{d {\bf v}_{i}}{dt} = {\bf F}_{i} \]

($\gamma =$ frictional drag coefficient; ${\bf v}_{i} =$ velocity of
$i$th monomer) with the further assumption that the inertial term
$m_{i}
\frac{d {\bf v}_{i}}{dt}$ is negligible:

\[ \frac{m_{i}}{\gamma} \frac{d {\bf v}_{i}}{dt} \approx 0 .\]

In the present work we have set $\gamma = 1$.  Hence

\[ \Delta {\bf R}_{i} = \Delta t \; {\bf v}_{i} = \Delta t \left(\frac{1}{\gamma}\right) {\bf F}_{i} = \Delta t \; {\bf F}_{i}.\]

Finally, the new unconstrained positions ${\bf R}_{i, unc} = {\bf
R}_{i, old} + \Delta {\bf R}_{i}$ are submitted to a standard SHAKE
routine~\cite{shake0} to enforce bond-length constraints.  For the
SHAKE routine, we used a tolerance of 0.01 times the bond-length
squared~\cite{shake1}.

We obtained our native structures by simulating homopolymer collapse,
with a penalty for contacts between neighboring residues along the
chain (more specifically, between residues $i$, $j$ satisfying $|i-j|
= 2$).

\section*{Results}

\begin{figure}
\begin{center}
\scalebox{0.8}{
\includegraphics[bb=61 0 480 254,clip=true]
{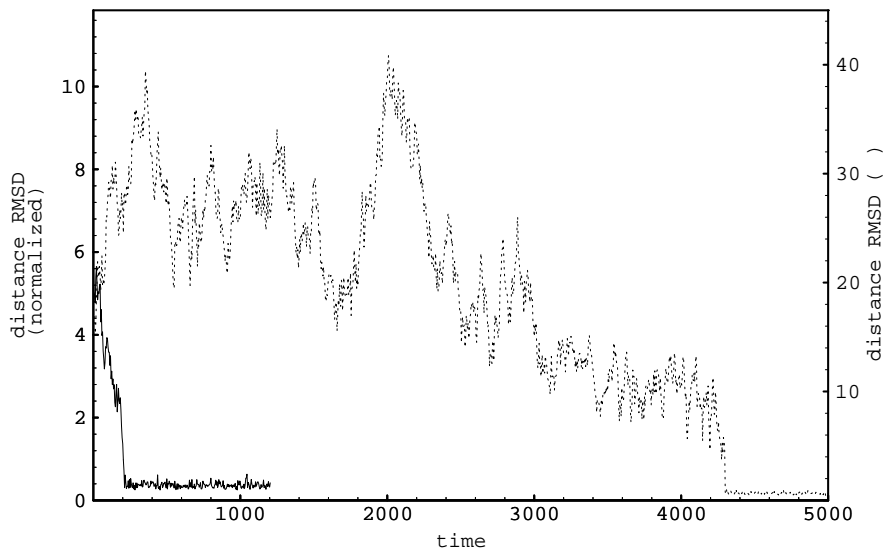}}
\end{center}
\renewcommand{\baselinestretch}{1}
\caption{
Typical folding trajectories for one 70-mer, using the isotropic
(solid line) and anisotropic (dashed line) models.  For both
trajectories $T = 0.6$.}
\label{typicalFoldingRmsd}
\end{figure}

The isotropic model described in the Model section above succeeded in
folding model proteins of up to length 70 (the longest we studied).  A
typical folding trajectory for a 70-mer is shown in
Figure~\ref{typicalFoldingRmsd}.

To study the nature of the unfolded $\leftrightarrow$ folded
transition, we focused on the behavior of one particular 30-mer
(s30.1).  This structure was obtained by homopolymer collapse at a low
temperature, followed by a very brief period of steepest-descent
energy minimization.  It contains 111 native contacts, and has a
native-state energy of -88.7.  Moreover, its radius of gyration is
1.77, which is close to maximally compact for a 30-mer.

\begin{figure}
\begin{center}
\includegraphics{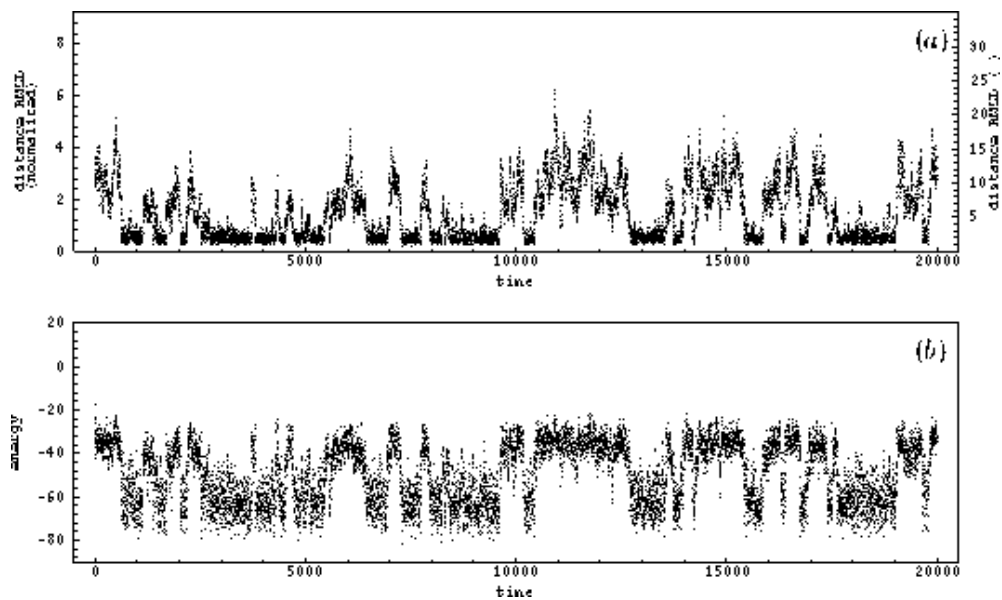}
\end{center}
\caption{Long folding trajectory at $ T = 0.64 \approx
T_f$, using the isotropic model.  The top graph ($a$)shows distance
RMSD vs. time (scale on the left is normalized, or bond-length, units;
the scale on the right is in angstroms).  The bottom graph ($b$) shows
energy vs. time.}
\label{isoLong}
\end{figure}

\begin{figure}
\begin{center}
\scalebox{0.8}{\includegraphics{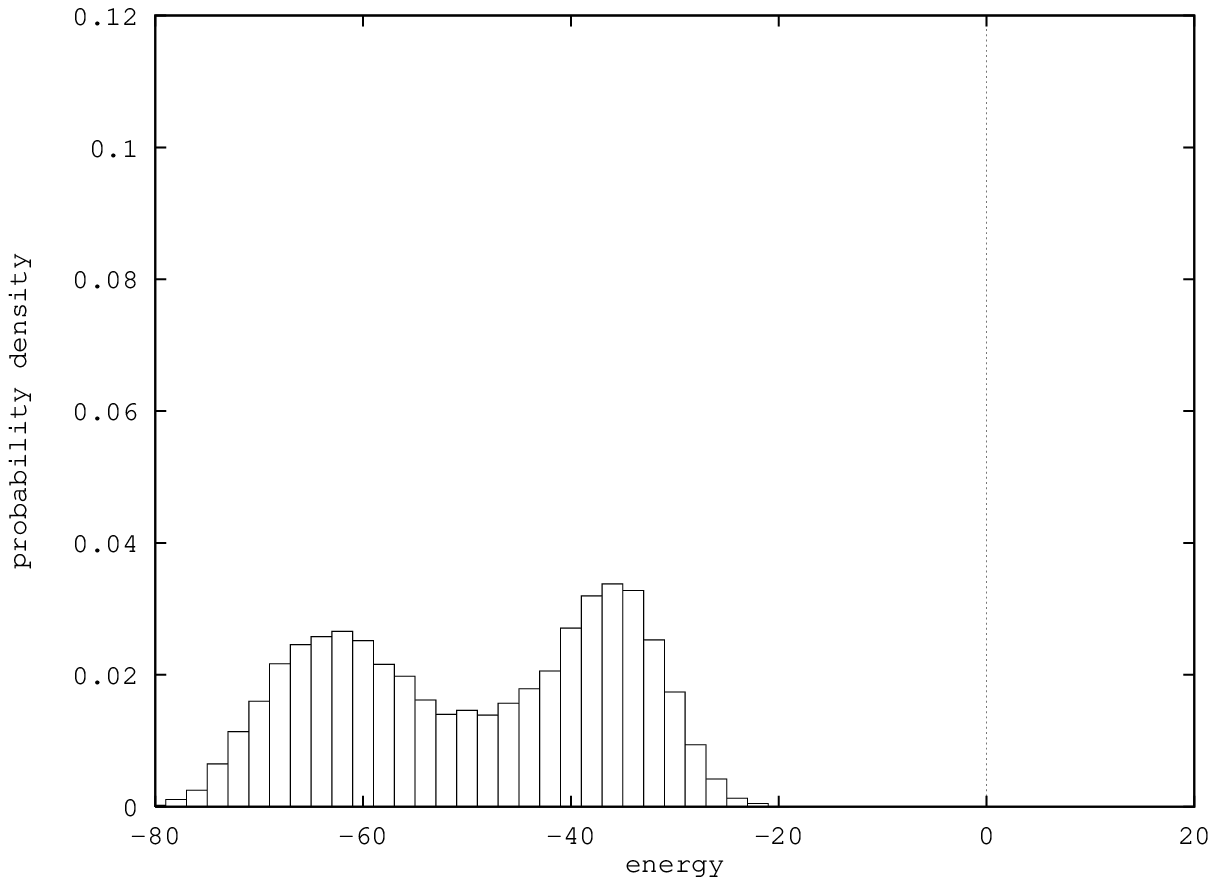}}
\end{center}
\renewcommand{\baselinestretch}{1}
\caption{
Distribution of energies for trajectory shown in
Figure~\ref{isoLong}.}
\label{isoHistoEnergy}
\end{figure}

\begin{figure}
\begin{center}
\scalebox{0.8}{\includegraphics{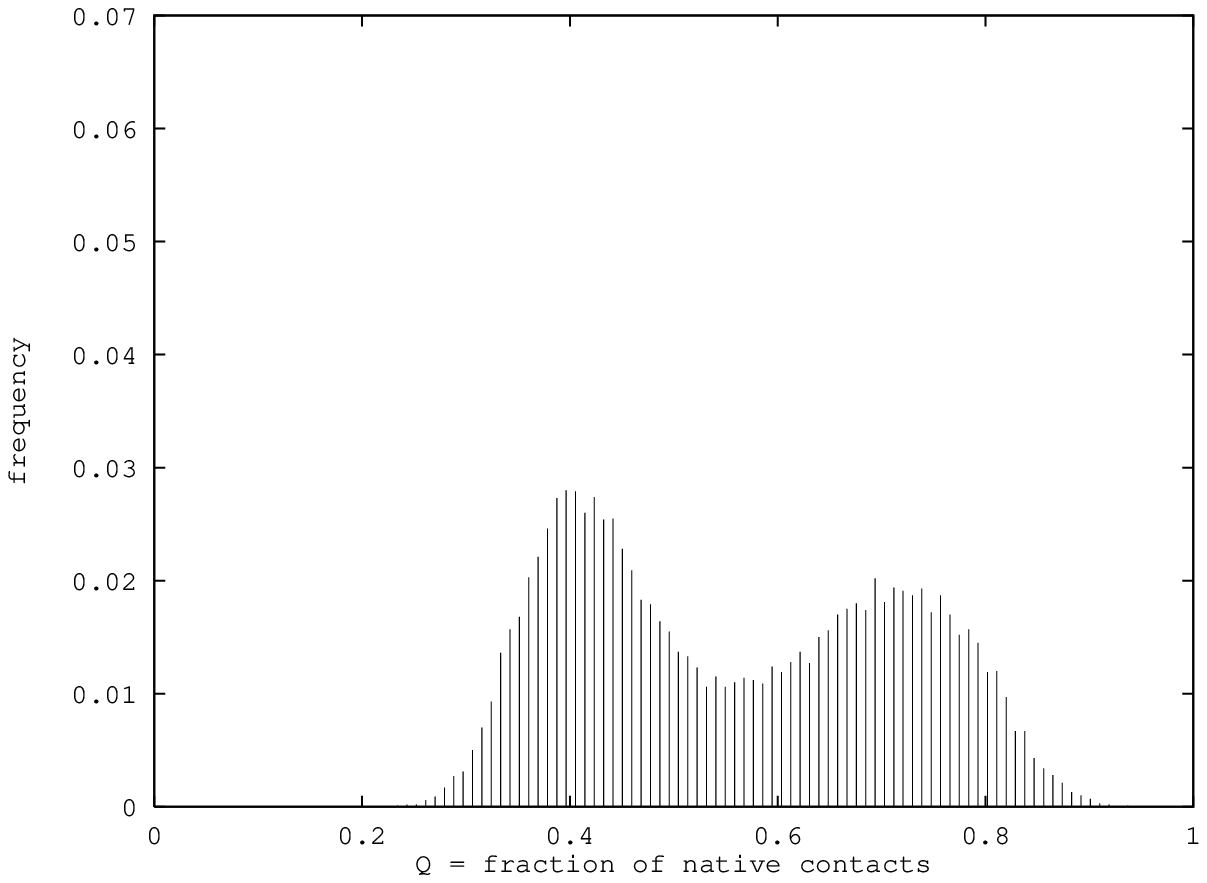}}
\end{center}
\renewcommand{\baselinestretch}{1}
\caption{
Frequencies of $Q$-values ($Q$ = fraction of native contacts) for
trajectory shown in Figures~\ref{isoLong}.}
\label{isoQTally}
\end{figure}

To gauge the degree of cooperativity in the transition between the
folded and the unfolded states, we first determined the transition
temperature for this structure to be $T_f \approx 0.64$.  We then
collected a long time series, at this temperature, of the distance
RMSD and the energy (Figure~\ref{isoLong}), as well as the fraction of
native contacts ($Q$, data not shown).  A histogram plot of the
enthalpies collected during this trajectory
(cf. Figure~\ref{isoHistoEnergy}) and the corresponding frequency
tallies (cf. Figure~\ref{isoQTally}) of the fraction of native
contacts ($Q$), show only a modest degree of cooperativity for the
isotropic model.

\begin{figure}
\subfigure[CI2]{
\label{65mersEnergyHisto:a}   
\begin{minipage}[b]{0.3\textwidth}
\begin{tabular}{c}
\includegraphics[width=\textwidth]{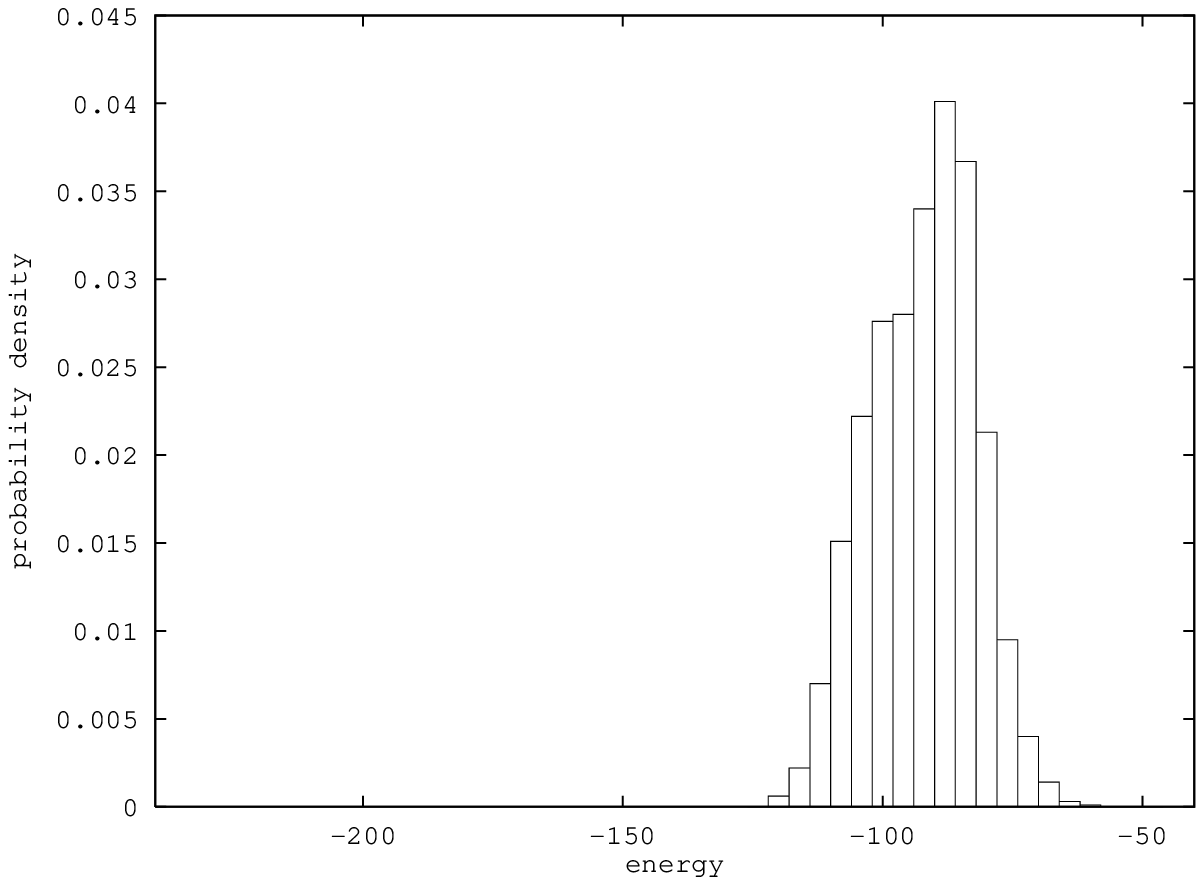}\\
\includegraphics[width=\textwidth]{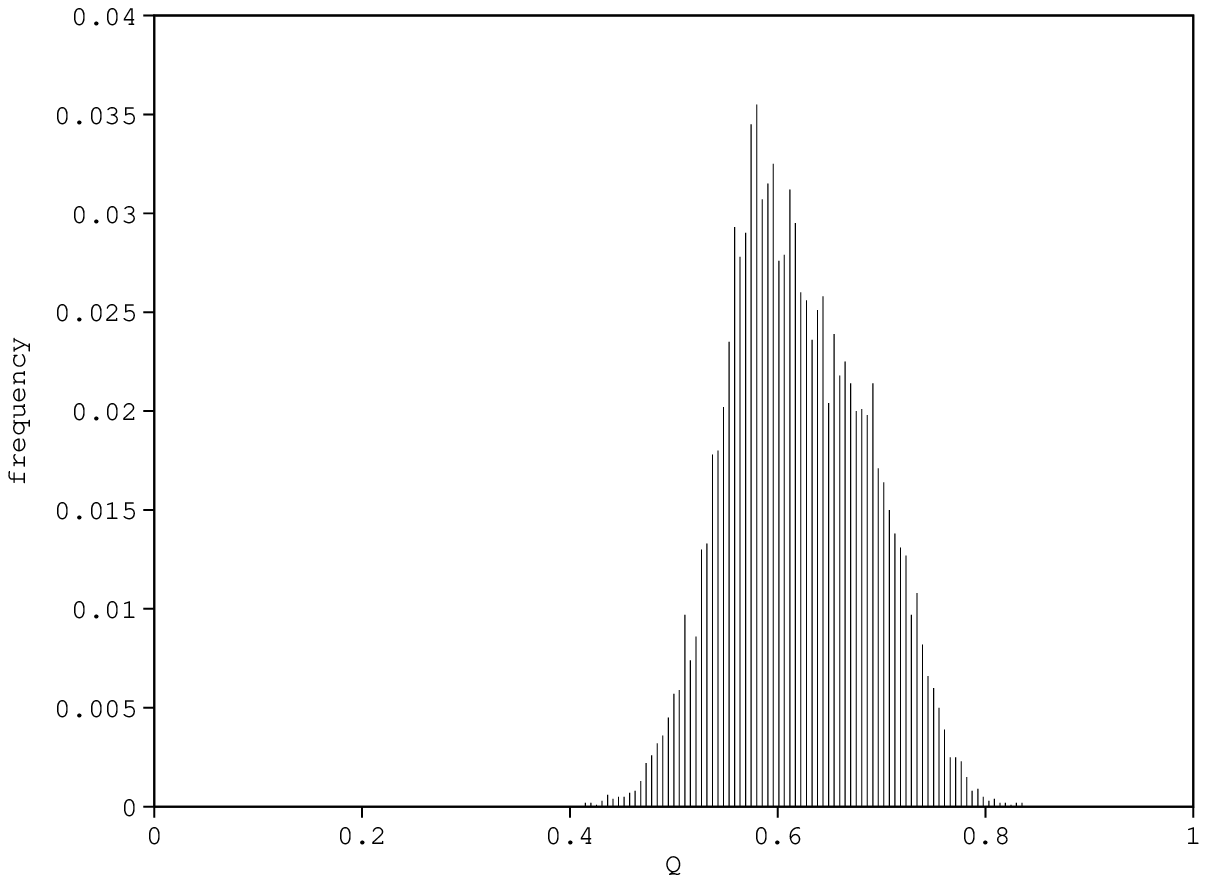}
\end{tabular}
\end{minipage}}%
\hfill%
\subfigure[s65.2]{
\label{65mersEnergyHisto:b}   
\begin{minipage}[b]{0.3\textwidth}
\begin{tabular}{c}
\includegraphics[width=\textwidth]{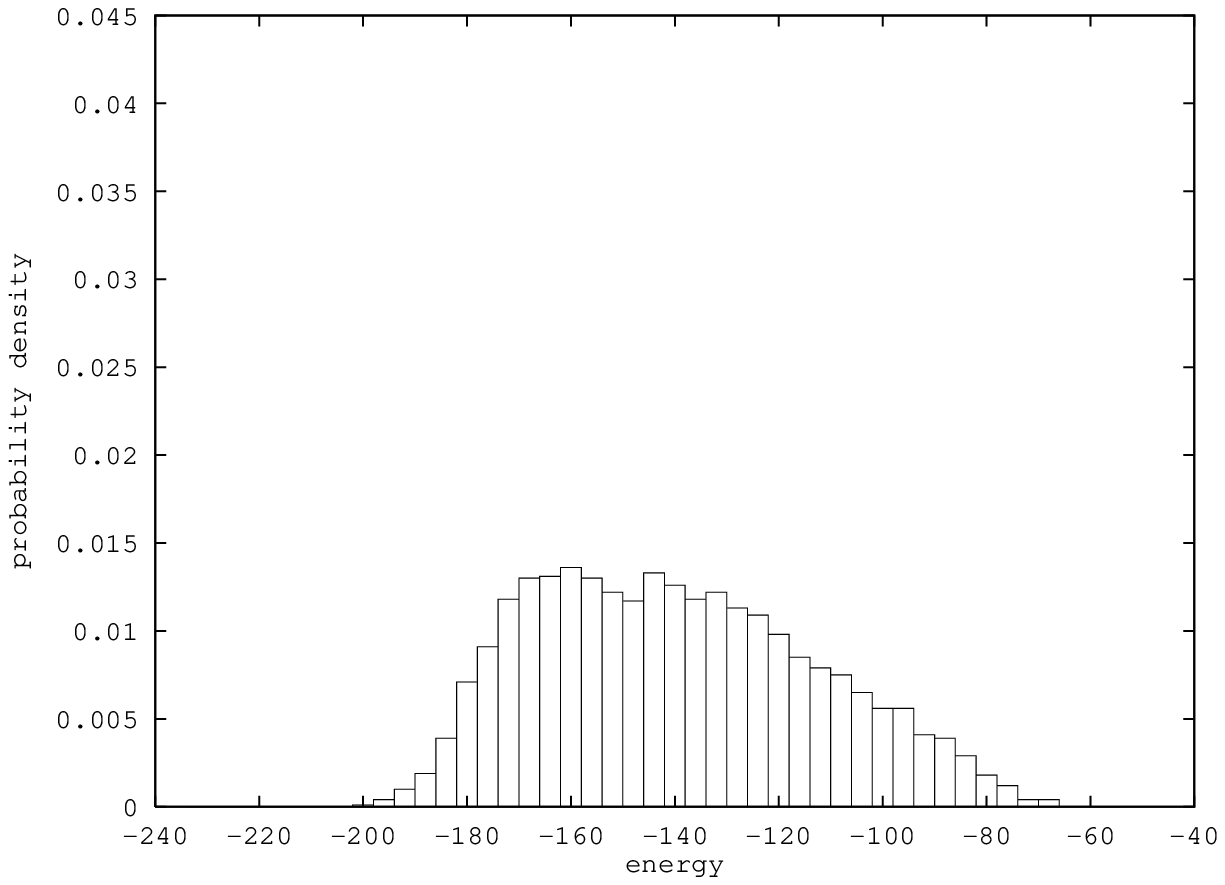}\\
\includegraphics[width=\textwidth]{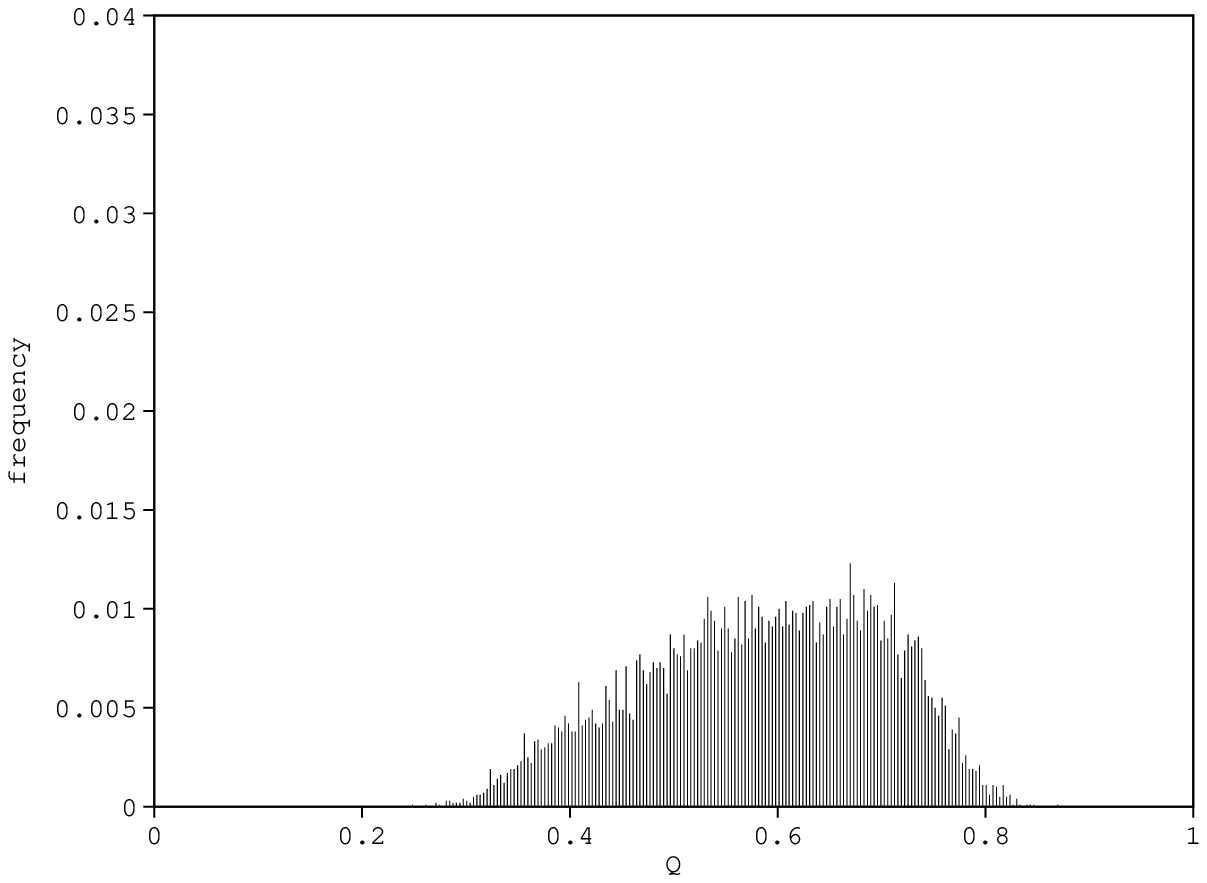}
\end{tabular}
\end{minipage}}%
\hfill%
\subfigure[s65.3]{
\label{65mersEnergyHisto:c}   
\begin{minipage}[b]{0.3\textwidth}
\centering
\begin{tabular}{c}
\includegraphics[width=\textwidth]{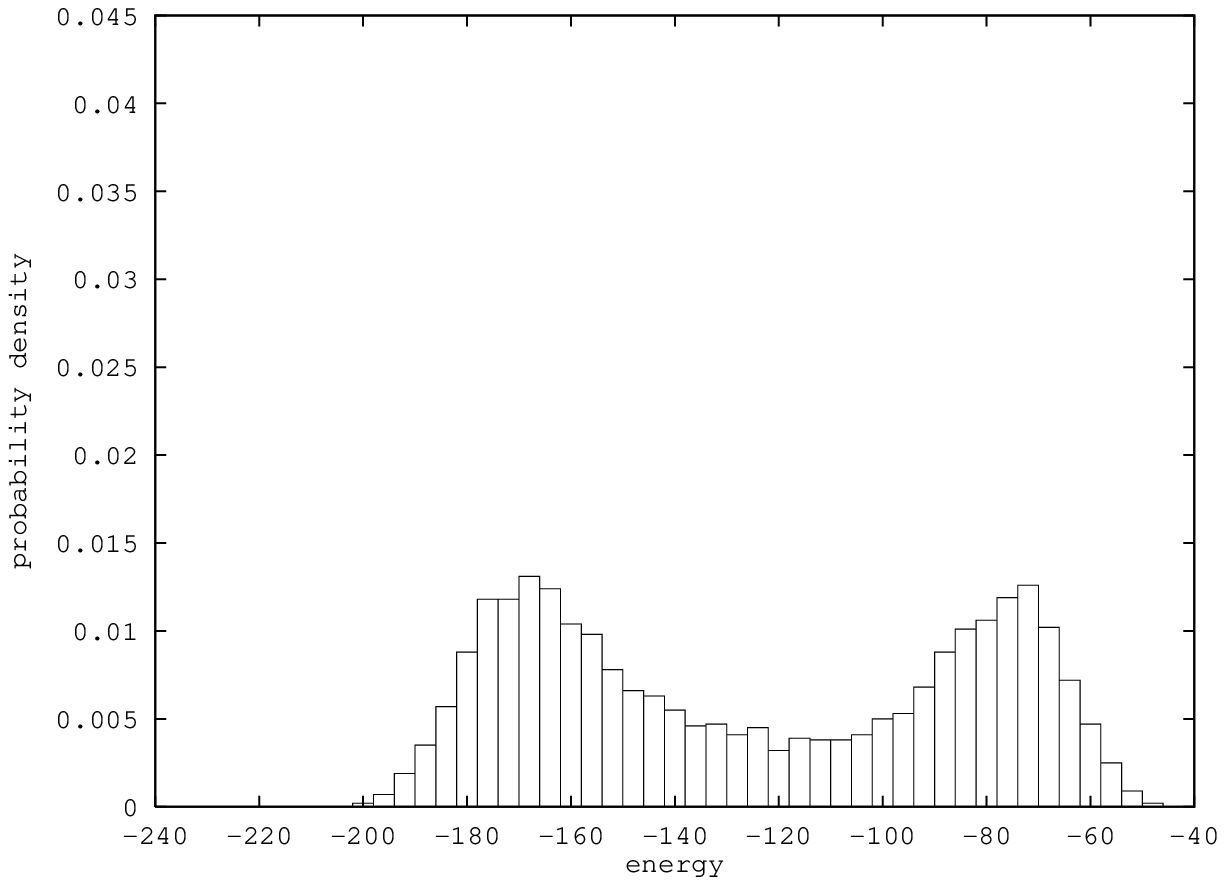}\\
\includegraphics[width=\textwidth]{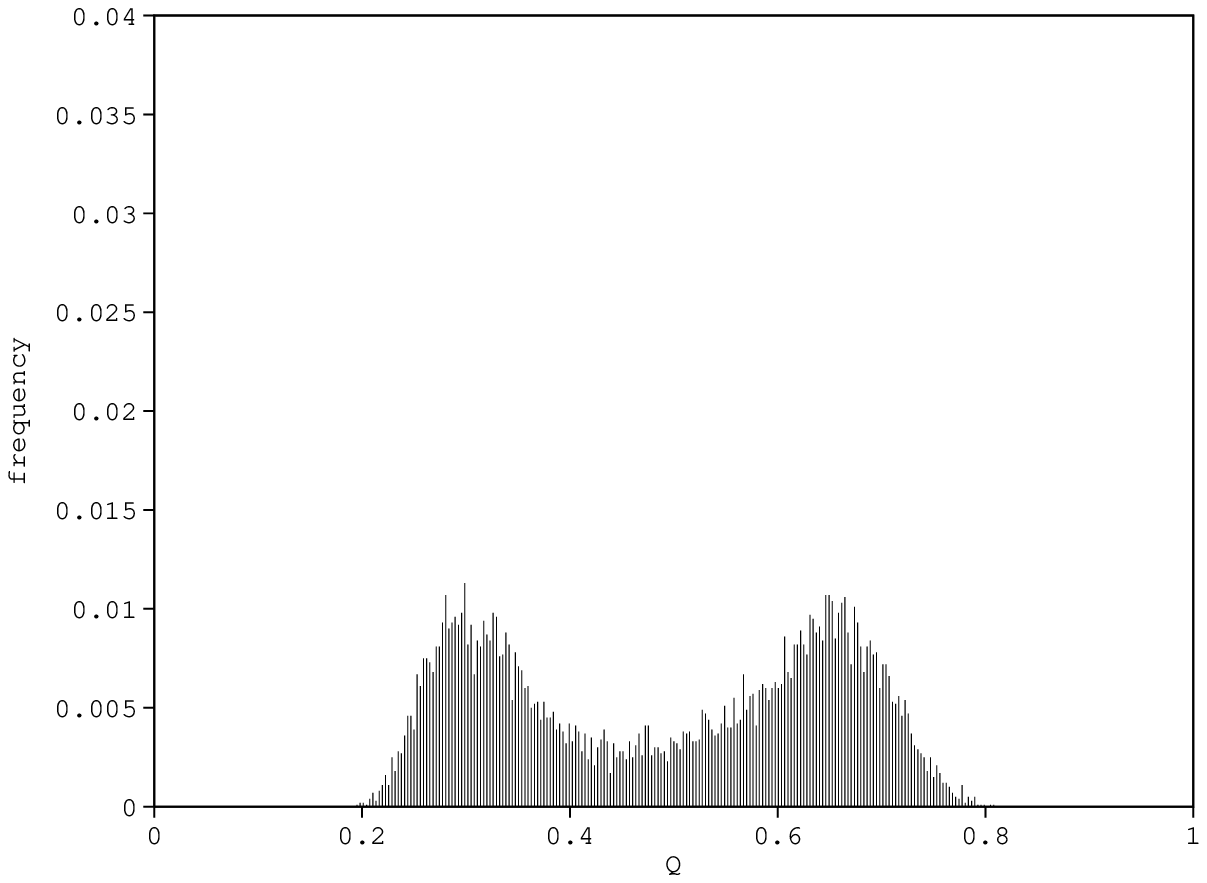}
\end{tabular}
\end{minipage}}
\caption{Energy distribution near $T_f$ for three 65-mers.}
\label{65mersEnergyHisto}
\end{figure}

To determine how much the above results depended on chain length, we
performed a similar analysis for three different 65-mers.  The first
of these was obtained from the $C_{\alpha}$ trace of chymotrypsin
inhibitor 2 (CI2).  The second and third (s65.2 and s65.3) were
obtained by homopolymer collapse at a low temperature, though for the
third one, the collapse was carried out for a longer period of time
than for the second one, resulting in a more compact structure.
Various relevant measurements for these three structures are given in
Table~\ref{comparisonChart65mers}.  Histograms for the energy
time-series for each structure (near their respective transition
temperatures) are shown in Figure~\ref{65mersEnergyHisto}.

\begin{table}[b]
\renewcommand{\baselinestretch}{1}
\begin{center}
\begin{tabular}{|r|ccc|}
\hline
structure & CI2 & s65.2 & s65.2 \\
\hline \hline
gyration radius & 2.88 & 2.63 & 2.32 \\
native-state energy & -49.5 & -233 & -241\\
approximate $T_f$ & 0.54 & 0.74 & 0.81 \\
number of native contacts & 188 & 306 & 328 \\
median $|i - j|$ for native contacts & 4.5 & 8 & 10 \\
\hline
\end{tabular}
\end{center}
\caption{Summary of physical characteristics of the three 65-mers
investigated.}
\label{comparisonChart65mers}
\end{table}

In the best case, that of structure s65.3, we see only a slightly
stronger cooperativity (as judged by the histograms' peak/trough
height ratios) than observed with the 30-mer s30.1 (compare with
Figure~\ref{isoHistoEnergy}).  It is interesting to see, however,
that the other two cases exhibit very little or no cooperativity at
all.  The CI2 structure is somewhat anomalous relative to the other
two 65-mers, as shown in Table~\ref{comparisonChart65mers}.  It has by
far the highest energy, not only for having the fewest native
contacts, but also for having a few strongly repulsive ones (the
monomers involved are slightly closer to each other than the
hard-sphere distance).  Moreover, most of its contacts are between
monomers that are relatively close to each other along the chain.  As
has been demonstrated with lattice models~\cite{ags95.0}, an abundance
of such ``local'' contacts weakens the cooperativity of the
transition.  It is suggestive that the median value, over all native
contacts, of their degree of ``locality'' $|j - i|$ (where $i$ and $j$
are the positions along the chain of the residues in the contact)
reflects the observed degree of cooperativity, at least within this
limited sample (cf. Table~\ref{comparisonChart65mers}).

\begin{figure}
\begin{minipage}[t]{\textwidth}
\begin{center}
\scalebox{0.8}{\includegraphics{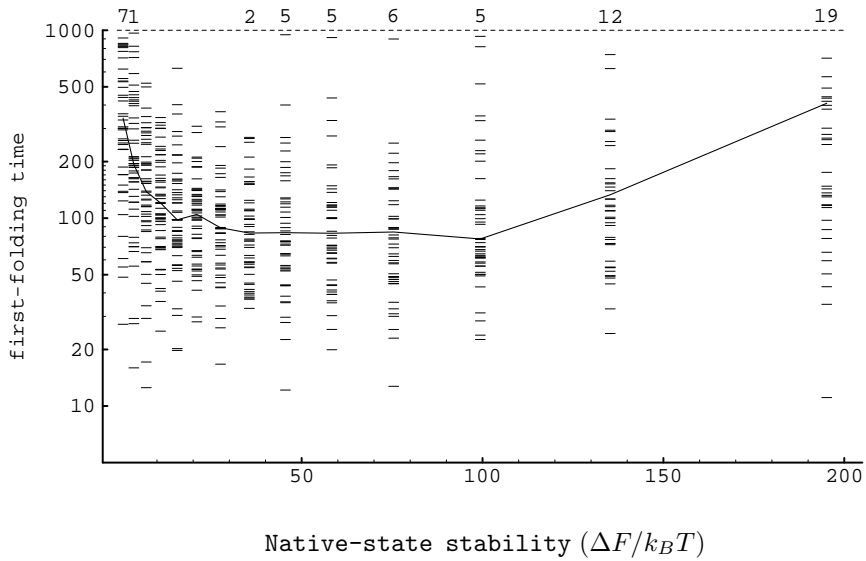}}
\end{center}
\end{minipage}
\begin{minipage}[t]{\textwidth}
\begin{picture}(13.5,0.4)
\footnotesize
\put(1.9,0){\makebox(10,0.4){{\tt Native-state stability} $(\Delta F/k_BT)$}}
\normalsize
\end{picture}
\end{minipage}
\renewcommand{\baselinestretch}{1}
\caption{
Dependence of first-folding time on native-state stability
($\frac{\Delta F}{k_BT}$) for isotropic model.  The curve shown (solid
line) is the sample median (50 observations per native-state
stability).  We used a cutoff of 1000 time units for these
simulations.  The integers shown above the dotted line at this cutoff
represent the numbers of runs, at the corresponding native-state
stabilities, that did not result in a folded structure within the cutoff
period (omitted when equal to zero).}
\label{isoMedianFolding}
\end{figure}

We also looked at the effect of native-state stability $\Delta F/T$ on
median first-passage time for the 30-mer s30.1.  (We estimated the
relation between the native-state stability and the inverse
temperature using the ``histogram method'' proposed by Ferrenberg and
Swendsen in~\cite{ferrenberg-swendsen88}.)  The results are shown in
Figure~\ref{isoMedianFolding} , where each data point represents the
median folding time of 50 folding simulations.  It is noteworthy that
the optimal median folding time is below 100 time units, but the
flattening of the curve and the data's noise preclude a satisfactory
determination of an optimal native-state stability for folding.

Spurred by the relatively weak degree of cooperativity shown by the
isotropic model, we devised the anisotropic model described in the
Model section above.  This model, too, was capable of folding proteins
of up to 70 residues in length (the longest we attempted).  A sample
folding trajectory for a 70-mer is shown in
Figure~\ref{typicalFoldingRmsd}.

\begin{figure}
\begin{center}
\includegraphics{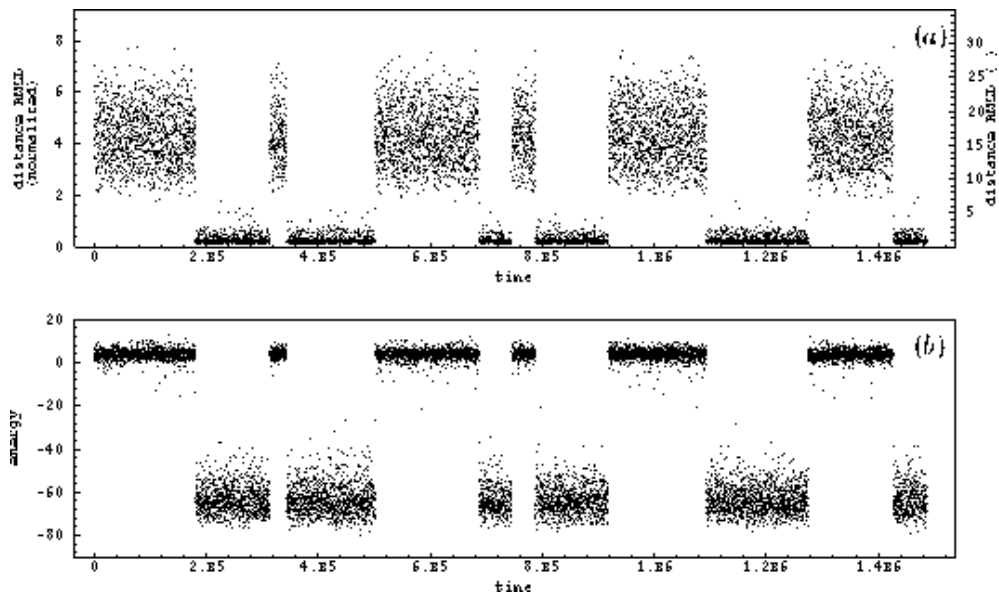}
\end{center}
\caption{Long folding trajectory at $ T = 0.725 \approx
T_f$, using the anisotropic model.  The top graph ($a$) shows distance
RMSD vs. time (scale on the left is normalized, or bond-length, units;
the scale on the right is in angstroms).  The bottom graph ($b$) shows
energy vs. time.}
\label{anisoLong}
\end{figure}

\begin{figure}
\begin{center}
\scalebox{0.8}{\includegraphics{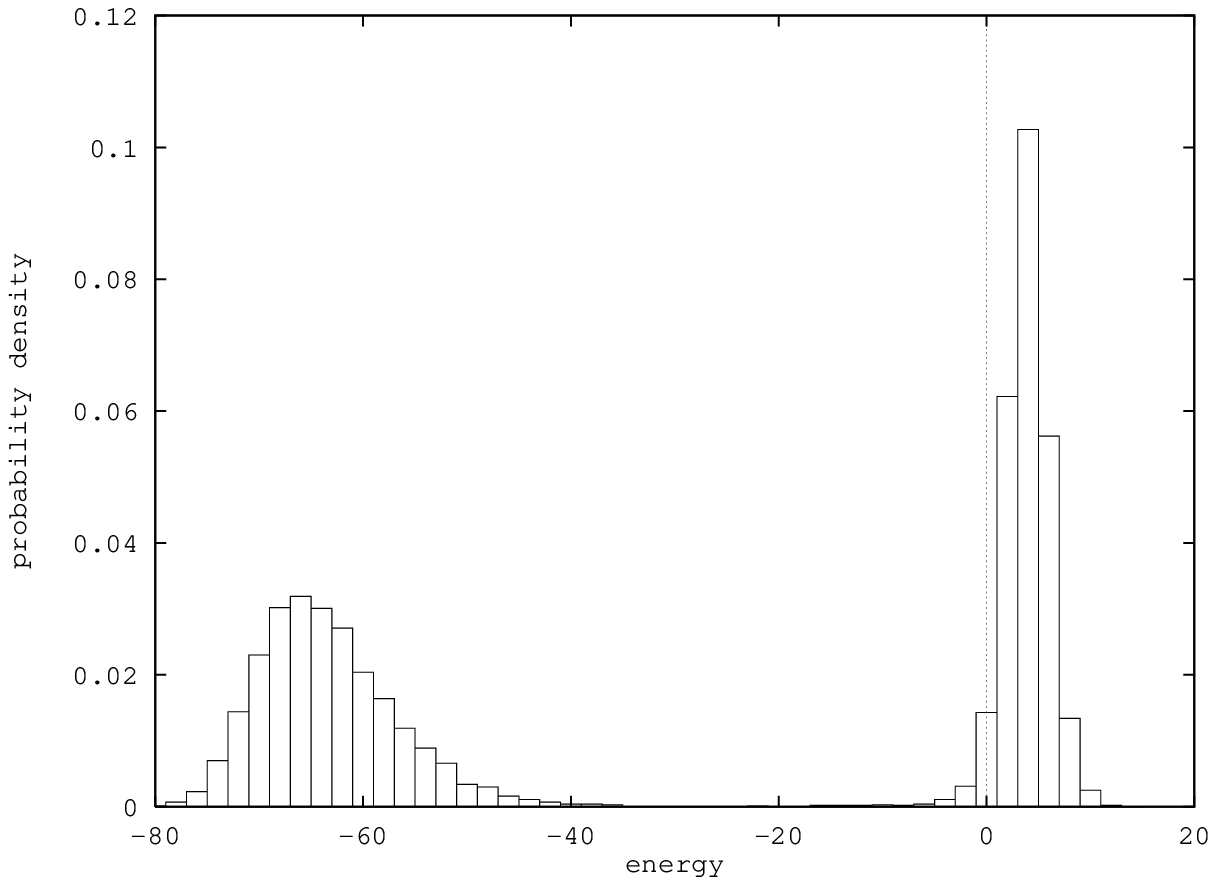}}
\end{center}
\renewcommand{\baselinestretch}{1}
\caption{
Distribution of energies for trajectory shown in
Figure~\ref{anisoLong}.}
\label{anisoHistoEnergy}
\end{figure}

\begin{figure}
\begin{center}
\scalebox{0.8}{\includegraphics{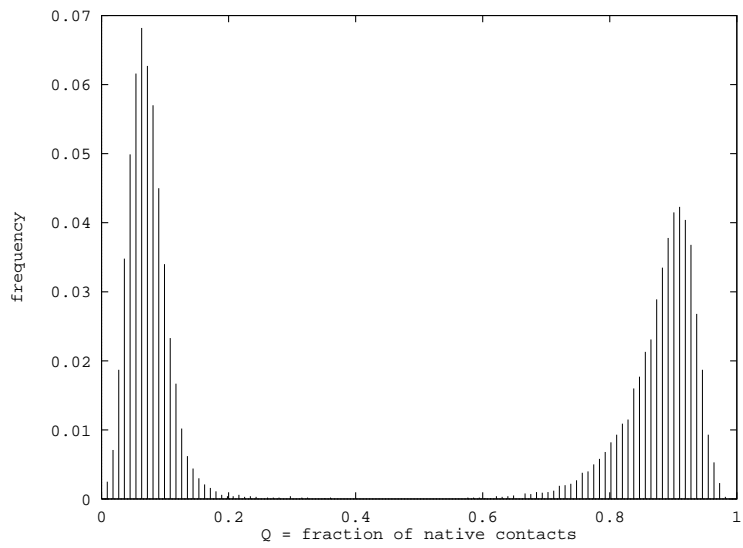}}
\end{center}
\renewcommand{\baselinestretch}{1}
\caption{
Frequencies of $Q$-values for trajectory shown in
Figure~\ref{anisoLong}.}
\label{anisoQTally}
\end{figure}

\begin{figure}
\begin{center}
\scalebox{0.8}{\includegraphics{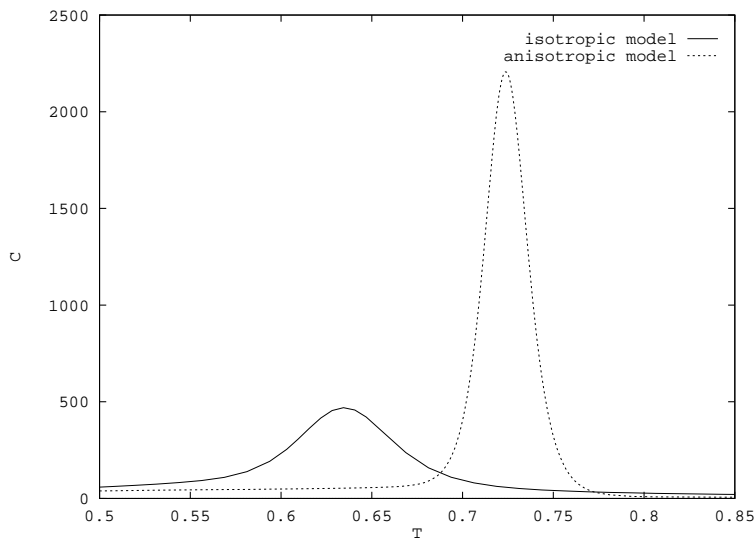}}
\end{center}
\renewcommand{\baselinestretch}{1}
\caption{
Heat capacity ($C$) vs. temperature for the isotropic and anisotropic
models.}
\label{cpVsT}
\end{figure}

We performed the same analyses as those performed with the isotropic
model, again using as our test structure the 30-mer s30.1.  A long
trajectory at $T_f \approx 0.725$ is shown in
Figure~\ref{anisoLong}\footnote{When comparing Figures~\ref{isoLong}
and~\ref{anisoLong}, it is important to be aware of the large
difference between the time (abscissa) scales for the isotropic and
the anisotropic models.}.  The corresponding histogram of enthalpies
and tally of Q-value frequencies (Figures~\ref{anisoHistoEnergy}
and~\ref{anisoQTally}) show a considerably greater degree of
cooperativity than observed with the isotropic model.  To further
quantify the strength of the first order transition, we plotted the
heat capacity $C \equiv dU/dT$ vs. $T$ (Figure~\ref{cpVsT}), computed
via the histogram method~\cite{ferrenberg-swendsen88} from the data
shown in Figures~\ref{isoLong} and~\ref{anisoLong}.  The maximum heat
capacity for the anisotropic model is almost 5 times greater than for
the isotropic one.  We estimated the transition barrier for each
model, from the data shown in Figures~\ref{isoQTally}
and~\ref{anisoQTally}; we obtained roughly 2 $k_BT$ and 6 $k_BT$, for
the isotropic and anisotropic models, respectively.

\begin{figure}
\begin{center}
\scalebox{0.8}{\includegraphics{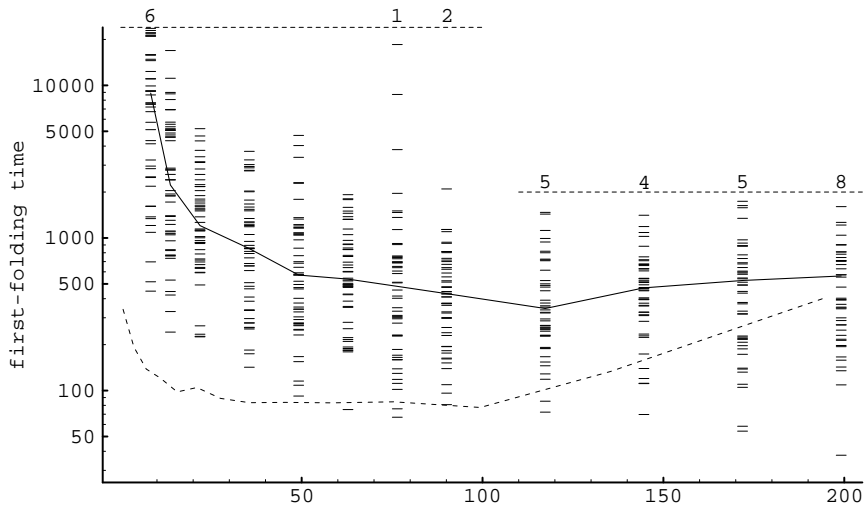}}
\end{center}
\renewcommand{\baselinestretch}{1}
\caption{
First passage times as a function of native-state stability
($\frac{\Delta F}{k_BT}$), for anisotropic model.  The upper curve
(solid line) is the sample median (50 observations per native-state
stability).  For these simulations, we used cutoffs of 24,000 and
2,000 time units, for the low and high native-state stability regions,
respectively, as indicated by the horizontal dotted lines.  The
integers shown above these dotted lines represent the numbers of runs,
at the corresponding native-state stabilities, that did not result in
a folded structure within the indicated cutoff period; omitted when
equal to zero.  The dotted curve appearing at the lower left is the
same median curve presented in Figure~\ref{isoMedianFolding}; it is
shown here for reference.}
\label{anisoMedianFolding}
\end{figure}

The effect of native-state stability on first-passage time is shown in
Figure~\ref{anisoMedianFolding}.  As was the case with the isotropic
model, the curve flattens in the high-stability region, which,
combined with the data's noise precludes an adequate estimation of the
native-state stability optimal for folding.  Taking into account these
caveats, we may estimate that the optimal median folding time in this
case is probably below 500 time units, certainly below 1000.

\begin{figure}
\begin{center}
\scalebox{0.8}{\includegraphics{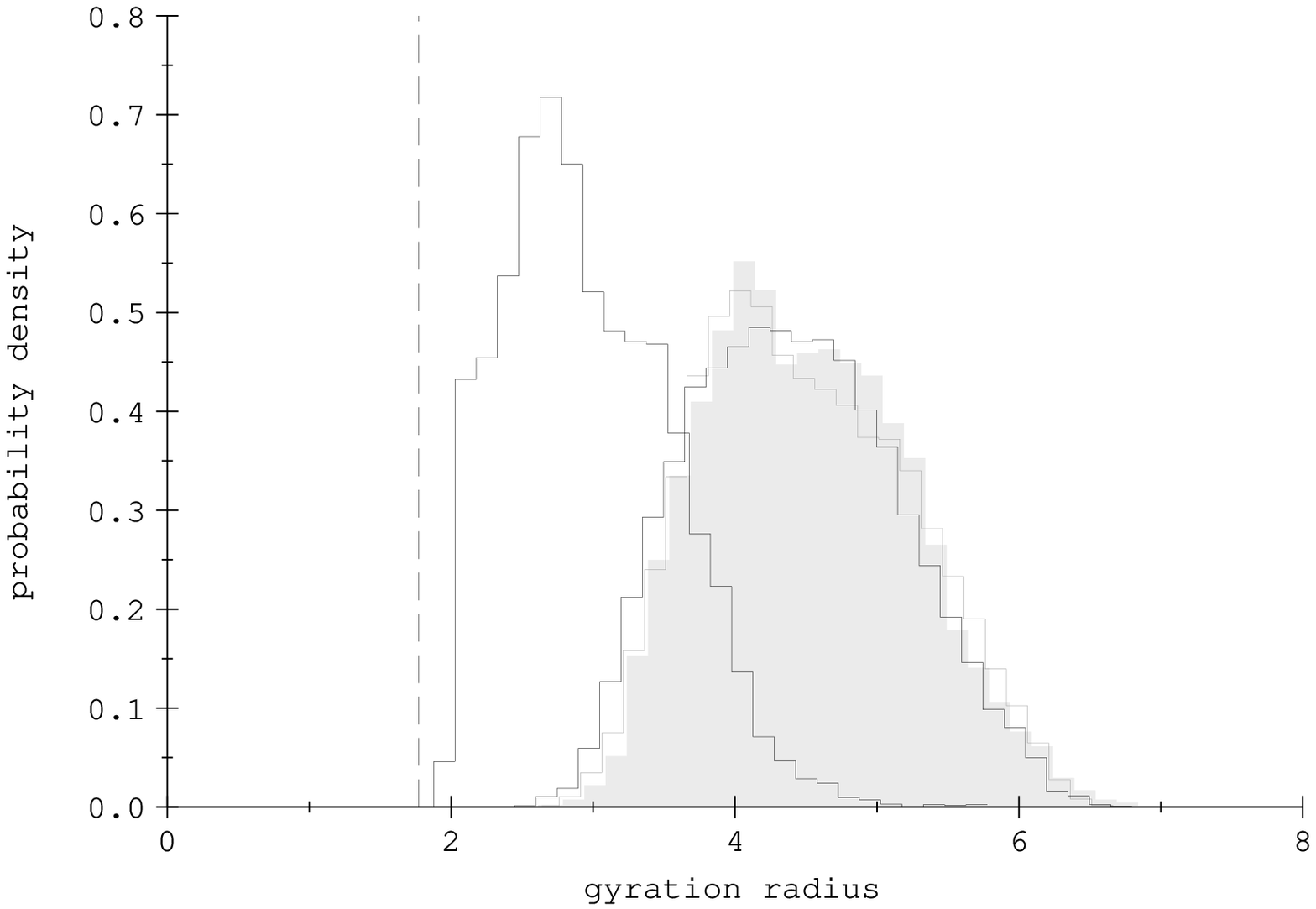}}
\end{center}
\renewcommand{\baselinestretch}{1}
\caption{
Distribution of radii of gyration.  The hump on the left corresponds
to the unfolded state under the isotropic model, at $T = 0.64$.  The
filled histogram on the right corresponds to the distribution
resulting from turning off all attractive potentials, at $T = 0.64$.
The (unfilled) histogram on the right with dark gray borders was
obtained under the same conditions as the filled histogram, but with
$T = 0.725$.  Finally, the remaining (black-edged, unfilled) histogram
on the right corresponds to the unfolded state under the anisotropic
model, at $T = 0.725$.  The dashed line on the left marks the radius
of gyration (1.77) of the native structure.  (For reference, bonds
have unit length, and the potential energy minima are at $r_0 =
1.5$.)}
\label{grAnalysis}
\end{figure}

We further investigated the nature of 30-mer s30.1's unfolded state
near $T_f$ under the isotropic and the anisotropic models.  As
Figure~\ref{grAnalysis} shows, the histogram for the radius of
gyration of the unfolded state in the anisotropic model is around 4,
almost indistinguishable from that of a simple self-avoiding chain
(i.e. one with no interaction between monomers other than hard-sphere
repulsion).  In contrast, with the isotropic model, the unfolded
state's mean radius of gyration is approximately 3.  Hence, with the
isotropic model, this 30-mer is never fully extended, at least at $T
\approx T_f$; instead, it remains in a somewhat compact globular state.  

For both models, at high native-state stabilities (low temperatures)
kinetic traps became increasingly common.  Each model, however,
exhibited traps of a characteristic type.  In the isotropic model, for
all the cases examined, the kinetically trapped structures consisted
of two well-folded domains of opposite chirality.  In the anisotropic
model chiral traps were never observed, as would be expected, since,
contrary to the isotropic model, the anisotropic one discriminates
between enantiomers.  Instead, almost all of the trapped structures we
observed with the anisotropic model were what we could call
``dumbbell'' traps: two perfectly folded halves that were, however,
ill-positioned with respect to one another.  The two exceptions we
found to this pattern were structures that were almost completely
folded except for short loops, towards the middle of each chain, that
could not attain their final buried positions due to the tightness of
the rest of the folded chain.  It is important to point out, however,
that for both models, kinetic traps were observed only at native-state
stabilities 3-15 times greater than those typical of real proteins.

\begin{figure}
\begin{center}
\scalebox{0.8}{\includegraphics{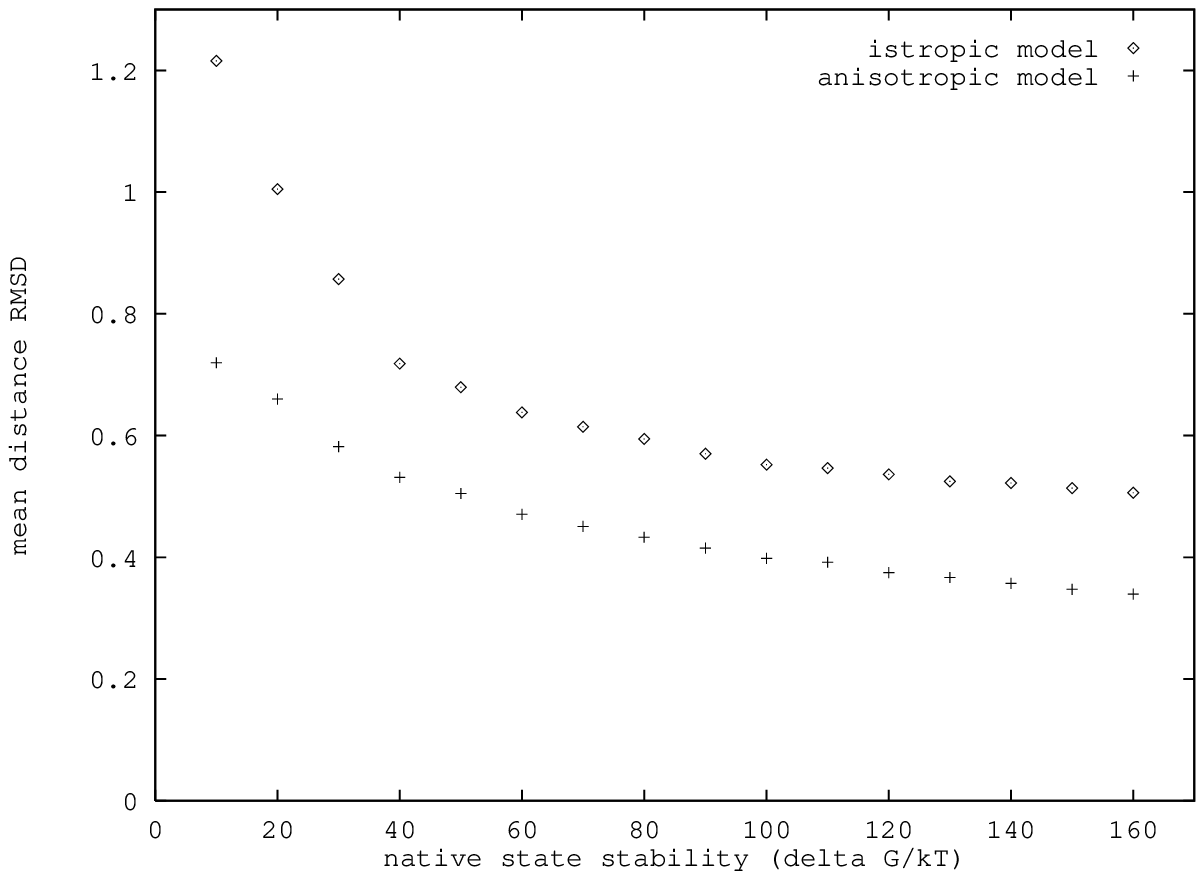}}
\end{center}
\renewcommand{\baselinestretch}{1}
\caption{Mean distance RMSD for the folded chain, as a function of the
native-state stability.}
\label{rmsdAnalysis}
\end{figure}

As would be expected, the two models also differ in the rigidity they
confer to the native state.  Figure~\ref{rmsdAnalysis} shows that, at
all native-state stabilities, the native state of the 30-mer s30.1 is
significantly more rigid with the anisotropic model.

\section*{Discussion}

We have presented two simple off-lattice models of protein folding,
which we have called the ``isotropic'' and ``anisotropic'' models,
respectively, in reference to the potential energy functions they use.
Although both models achieve the primary requirement of folding model
proteins (up to length 70, at least) to their native states, they show
several substantive differences.  These are summarized in
Table~\ref{comparisonChart}.

\begin{table}
\renewcommand{\baselinestretch}{1}
\begin{tabular}{|r|c|c|}
\hline
model & {\bf isotropic} & {\bf anisotropic}\\
\hline \hline
first-order transition & weak & strong \\
\hline
optimal median folding time ($t_{\mbox{\scriptsize opt}}$)& (70) & (350) \\
\hline
(median folding time at $T_f$) $\div\;t_{\mbox{\scriptsize opt}}$& (5) & $(10^2
- 10^3)$\\
\hline
gyration radius of unfolded state:
mean and (sd)& 3.1 (0.55) & 4.4 (0.73) \\
\hline
rigidity of native state & low & high \\
\hline
type of kinetic traps & chiral & dumbbell\\
\hline
\end{tabular}
\caption{Summary of functional differences in the folding of one
30-mer between the isotropic and
anisotropic models.}
\label{comparisonChart}
\end{table}

\begin{figure}
\begin{center}
\scalebox{0.8}{\includegraphics{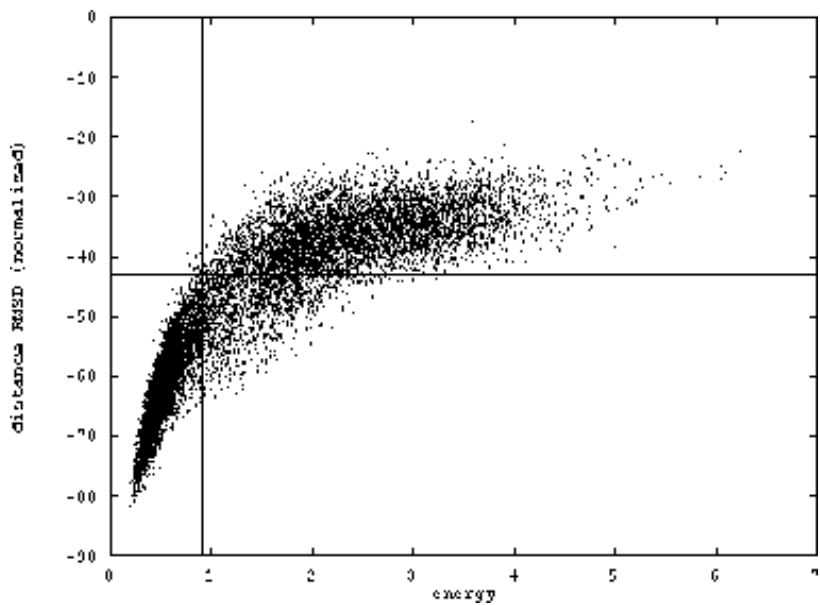}}
\end{center}
\renewcommand{\baselinestretch}{1}
\caption{
Energy vs. Distance RMSD plot for the data shown in
Figure~\ref{isoLong}.}
\label{isoEnergyVsRmsd}
\end{figure}

Our analysis of the folding transition for the isotropic model showed
that it was capable of delivering a modest, but detectable, degree of
cooperativity.  From the data shown in Figure~\ref{isoQTally}, we
estimate a barrier of 1-2 $k_BT$ at $T\approx T_f$.  Guo and
Brooks~\cite{guo-brooks96} have obtained very similar results in their
study of the off-lattice model first proposed
in~\cite{honeycutt-thirumalai90,honeycutt-thirumalai92}.  This
relatively weak first order transition delivered by the isotropic
model was the primary motivation behind our development of the
anisotropic one.  The plot of energy vs. distance RMSD at $T \approx
T_f$ for the isotropic model (Figure~\ref{isoEnergyVsRmsd}) suggested
to us that its spherically symmetrical potential energy function
resulted in the relative stabilization of a sizeable population of
states (namely, those in the region defined by a distance RMSD $>$ 0.9
bond lengths $\approx$ 3~\AA, and energy $<$ -43), that were unrelated
to the native one.  This, in turn, would promote a non-cooperative
component to the folding mechanism: the gradual rearrangement of a
partially collapsed unfolded state.

\begin{figure}
\begin{center}
\scalebox{0.8}{\includegraphics{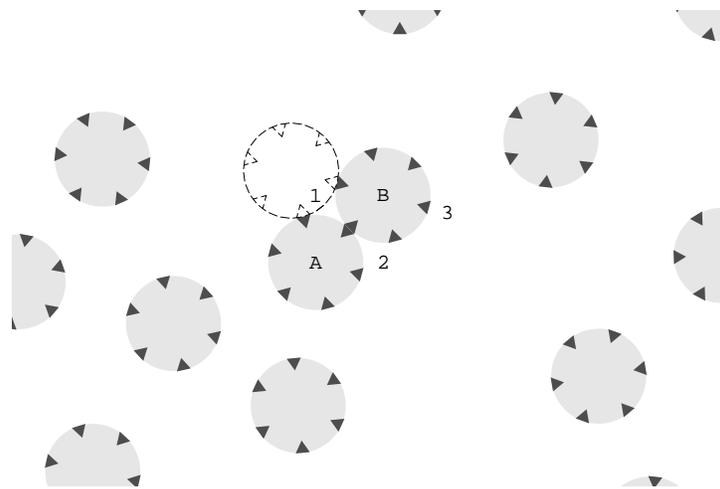}}
\end{center}
\renewcommand{\baselinestretch}{1}
\caption{Circular particles with hexagonal arrangement of interaction
sites.  The formation of a contact between particles A and B
simultaneously produces composite interaction sites 1 and 2.
}
\label{stericNucleation}
\end{figure}

We then hypothesized that rigidly asymmetric potentials would lead to
cooperative folding.  To illustrate the reasoning behind this
hypothesis, we propose a simple example in two dimensions.  Consider
first a system of identical circular particles (see
Figure~\ref{stericNucleation}) performing Brownian motion within a
bounded 2-dimensional space.  Suppose that these particles have 6
hexagonally arranged ``interaction sites,'' such that any two such
sites on different particles attract each other.  Let the change in
energy upon formation of one such contact be $\Delta U < 0$, and the
corresponding change in entropy (upon fixing of one particle relative
to the other) be $\Delta S < 0$.  Therefore, starting from a
``monoatomic'' state (i.e. all particles unattached) at temperature
$T$, the formation of the first pairwise contact entails a free-energy
change $\Delta F = \Delta U - T \Delta S$.  However, the very
formation of this first contact creates two new ``composite''
interaction sites (labeled 1 and 2 in
Figure~\ref{stericNucleation}), each consisting of two simple sites.
The crucial point is that now the arrival of a third monomer at one of
these new composite sites entails a free-energy change of $2\Delta U -
T \Delta S < 2\Delta F$.  Therefore, at any temperature, it is
thermodynamically more favorable to form two new contacts by adding a
third monomer to the dimer at one of the new composite sites, than by
bringing together two separate pairs of monomers (as would happen if
monomers attached at sites 3 and 4, for example).  As a consequence,
in this system, the fixed geometrical arrangement of a discrete set of
attachment sites on each monomer to cooperative aggregation.  We use
the term {\em steric nucleation} to refer to this interplay between
the thermodynamics of contact-cluster formation and the geometrical
arrangement of the interaction sites on each monomer.

Now, we consider a hypothetical protein.  Let us suppose that its
native state includes a ``triangular'' set of contacts between
monomers $i$ and $j$, $j$ and $k$, and $i$ and $k$.  (We will refer to
this as a ``3-cluster,'' short for ``3-contact cluster'').
Furthermore, suppose that, like in the 2D example above, the very
formation of any one of these three contacts orients the two
participating monomers in such a way that the subsequent arrival of
the third monomer results in the simultaneous formation of the
cluster's remaining two contacts.  Then, reasoning as before, we
expect that the formation of this contact cluster will be cooperative.
Taking this reasoning one step further, we see that if a native
structure consisted of an interconnected network of such $n$-clusters
($n \geq 3$), then the entire folding event would be cooperative.

It is plausible that some form of steric nucleation contributes to the
cooperativity observed in the folding of real proteins.  Since amino
acids are completely asymmetrical molecules, we expect that, in
general, the change in energy resulting from bringing two of them
together will depend not only on the distance between them, but also
on their relative orientations.  This would mean that the energy
landscape for the interaction between two amino acids would feature a
few discrete minima.  In the language used above, protein optimization
may involve assembling $n$-clusters of spatially complementary
contacts into an interconnected network spanning the entire native
structure.  To be sure, it is unlikely that all, or even most, of the
intramolecular interactions contributing to the energy of folding of a
real protein are as narrowly specific as the active contacts built
into our anisotropic model.  A more realistic possibility is that, in
a real protein, only a fraction of these interactions have the steric
stringency of the anisotropic model, but that, nonetheless, these
alone are sufficient to render the folding cooperative.  This
hypothesis immediately proposes the investigation of a hybrid model,
in which some contacts are sterically restricted, while others are
not.  More precisely, we may ask, what is the relation between the
fraction and/or strength of contacts that are sterically restricted,
and the degree of cooperativity of the folding transition?  And can a
small subset of such contacts serve as a {\em specific} nucleus for
the folding of the whole chain?  Another interesting study would be of
the structures of proteins homologous to one of those for which a
folding nucleus has been experimentally identified, to see if the
residues corresponding to nucleus sites exhibit a greater degree of
spatial overlap across the various homologs than would be otherwise
expected.  These are investigations we are currently pursuing.

We devised the anisotropic model as a simple modification of the
isotropic one that would be capable of specifically testing the steric
nucleation hypothesis presented above, within the context of
off-lattice model-protein folding.  We believe it has served this
purpose well.  However, in the name of expediency and computational
efficiency, for its implementation we chose to keep all the sets of
vectors ${{\bf\hat{u}}_{ij}}$ (cf. Model section) at fixed, concordant
orientations throughout the simulation.  It is likely that the greater
first-folding times of the anisotropic model are due, at least in
part, to this feature of our implementation.  Indeed, the latter
implies that any rotation of the native structure, or more
importantly, of any small fragment thereof, will be destabilizing.
This, in turn, implies a higher kinetic barrier for the anisotropic
model than would be expected for a model that did not impose a fixed
spatial orientation on the native state.  A more generally applicable
implementation of the anisotropic model than the one we chose would
have allowed all the vectors ${\bf\hat{u}}_{ij}$, for any given
monomer $i$, to rotate together as a rigid unit, and independently of
any other set ${\bf
\hat{u}}_{i\,'\,j}$.  Such a model would still support steric
nucleation, but it would also allow $n$-clusters to form in several
spatial orientations that would, in general, would be all different
from each other.  Therefore, even though such a model would give each
pair of monomers 3 more degrees of freedom, thereby reducing the
probability of nucleus formation, those nuclei that did form would be
much less fragile than those in our version here.  Hence, it is
possible that, despite the increase in the size of the chain's
conformational space, such a generalized model would result in faster
folding times.

Incidentally, it is worth remarking that for a 30-mer on a cubic
lattice, and still using a Go-type model, the optimal median folding
time is about 27000 Monte Carlo steps~\cite{gutin-chainLength}; upon
dividing this figure by the chain length (to correct for the
difference in the counting of time between the Monte Carlo methods and
the off-lattice methods presented here), we get 900 time units, or
roughly 1 order of magnitude slower than for the optimal median
folding time isotropic model.  This slowness of the lattice relative
to the isotropic model probably reflects the lattice's much more
limited geometry.

It is clear from the energy trajectory in Figure~\ref{isoLong} (or from
Figure~\ref{isoHistoEnergy}) that with the isotropic model, a
significant fraction of the native contacts are present in the
unfolded state (for both models, the energy of a perfectly stretched
conformation is zero).  Moreover, with the isotropic model the
unfolded state is significantly more compact, as measured by the
radius of gyration, than it is with the anisotropic model.  With the
anisotropic model, on the other hand, the energy of the unfolded state
is slightly above zero, consistent with very few native contacts (and
a few unfavorable hard-ball repulsions).  Indeed, we found that, at $T
\approx T_f$, with the isotropic model the average value for the
fraction $Q$ of native for the unfolded state was between 0.4 and 0.5,
whereas with the anisotropic model this figure was below 0.1.  (See
Figures~\ref{isoQTally} and~\ref{anisoQTally}.)  Moreover, for the
latter, $Q-$values between 0.3 and 0.6 are almost never observed.
These data suggest that, in the anisotropic model, the transition
state for the structure under study consists of conformation(s) with
about $30\%$ of the native contacts.  (Further study will reveal how
specific this $30\%$ needs to be to ensure folding.)  In contrast, the
corresponding figures for the isotropic model suggest that when
pairwise distance is the only requirement for contact formation, there
are many unproductive ways of making contacts
(Figure~\ref{isoQTally}).  Finally, in relation to the foregoing, we
should note that the anisotropic model seems to recapture one of the
features of the lattice that was lost by the isotropic model, namely a
very small number of ways in which $n$-clusters may form.  While the
lattice features anisotropy inherently, it must be explicitly
introduced into off-lattice models.

We should also note that it is possible that the relative compactness
of the unfolded state in the isotropic model contributes to its
greater folding speed.  True, with the isotropic model, there appears
to be a greater likelihood that the chain will spend time sampling
unproductive conformations of relatively low energy.  However, even at
modest native-state stabilities, the isotropic model folds within a
mere one hundred time units.  Thus, it is possible that the net effect
of compactization is to enhance folding rate.  Further investigation
is necessary to sharpen our understanding of this matter.

Folding was generally faster with the isotropic model than with the
inosotropic one at all native-state stabilities studied, although the
difference becomes negligible (to within the data's noise) at the
highest native-state stabilities.  As just alluded to, with the
anisotropic model, the chain is not as likely to linger in misfolded
states of relatively low energy, as it is with the isotropic model.
On the other hand, with the anisotropic model, contact formation will
certainly be more infrequent than with the isotropic model, since,
with the former, and not with the latter, contact formation requires
residues to approach each other in a precise orientation.  It appears
(Figure~\ref{anisoMedianFolding}) that the latter effect dominates the
kinetics of the anisotropic model, at least for most of the range of
native-state stabilities studied.  Interestingly, with both models,
the median folding time curves plateau at high native-state
stabilities (Figures~\ref{isoMedianFolding}
and~\ref{anisoMedianFolding}).  We hypothesize that this is an
artifact of Go's prescription.  Indeed, in a sequence model
energetically favorable non-native contacts are possible, which would
result in a faster increase in median folding time as a function of
increasing native-state stabilities than with the models presented
here.

It is at temperatures close to $T_f$, however, that the two models
presented here differ most dramatically.  For the isotropic model, the
ratio of the folding time at the transition temperature $T_f$ and the
optimal folding time $t_{\mbox{\scriptsize opt}}$ is roughly 5.  For
the anisotropic model, we cannot give a similarly precise figure, due
to our inability to collect adequate statistics for this model at
$T_f$.  However, it is clear from our experience so far with this
model (typified by the time course in Figure~\ref{anisoLong}) that
this ratio is well above 100, and probably lies somewhere between 300
and 1000.  This difference can be seen as indicative of the greater
role of nucleation in the anisotropic model, since the rate of nucleus
formation is very sensitive to high temperatures.

In this connection, it is interesting to note a very remarkable
property of real proteins, namely that their folding rate decreases by
up to three orders of magnitude as the stability of the native state
is reduced to
zero~\cite{jackson-fersht91.1,khora-roder93,itzhaki-otzen-fersht95}.
In light of this experimental fact, it is encouraging to see that the
anisotropic model produces a commensurate retardation of folding at
zero native-state stability.  Our enthusiasm is tempered, however,
upon noting that real proteins achieve this retardation over a range
of native-state stabilities of only 10-20 $k_BT$, while with our
anisotropic model it occurs over a 10-fold greater range.  Although we
expect to see a significant change in folding kinetics once we
generalize the anisotropic model (to allow freely rotating contacts),
it is not immediately clear how this generalization will affect either
the range of folding rates, nor the corresponding range of
native-state stabilities.

\section*{Acknowledgments}

We would like to thank Victor Abkevich and Michael Morrissey for many
fruitful discussions; and Zhuyan Guo and Charles L. Brooks, for
sharing their observations with us prior to publication.

\clearpage
\bibliography{main}
\bibliographystyle{aip}
\end{document}